\documentclass[letter]{aa}
\usepackage{multirow}
\usepackage{amsmath}
\usepackage{graphicx,subfig} 
\usepackage{lscape}
\usepackage{longtable}
\usepackage{txfonts}
\usepackage{color}
\usepackage{booktabs}
\usepackage{soul}
\usepackage{rotating}
\usepackage[nolist,nohyperlinks]{acronym}

\makeatletter
\renewcommand*\aa@pageof{, page \thepage{} of \pageref*{LastPage}}
\usepackage{hyperref}
\hypersetup{colorlinks=true,allcolors=[rgb]{0,0,0.8}}
\hypersetup{
  colorlinks=true,
  allcolors=[rgb]{0,0,0.8},
  pdftitle={eROSITA All-Sky Survey Cluster Catalog},
  pdfauthor={Bulbul et al.},
  }

\newcommand{\xmm}{\textit{XMM-Newton}\xspace}

\newcommand{\esass}{\texttt{eSASS}\xspace}

\acrodef{cxb}[CXB]{cosmic X-ray background}
\acrodef{nxb}[NXB]{non-X-ray background}
\acrodef{lhb}[LHB]{Local Hot Bubble}
\acrodef{gh}[GH]{Galactic Halo}
\acrodef{oot}[OoT]{Out-of-Time}
\acrodef{fwc}[FWC]{filter-wheel-closed}
\acrodef{cr}[CR]{cosmic ray}
\acrodef{igrm}[IGrM]{intragroup medium}
\acrodef{arf}[ARF]{ancillary response file}
\acrodef{rmf}[RMF]{redistribution matrix file}
\acrodef{sz}[SZ]{Sunyaev-Zeldovich}

\usepackage{natbib,twoopt}
\usepackage{xcolor}

\makeatletter
\newcommandtwoopt{\citeads}[3][][]{\href{http://adsabs.harvard.edu/abs/#3}%
    {\def\hyper@linkstart##1##2{}%
     \let\hyper@linkend\@empty\citealp[#1][#2]{#3}}}
  \newcommandtwoopt{\citepads}[3][][]{\href{http://adsabs.harvard.edu/abs/#3}%
    {\def\hyper@linkstart##1##2{}%
     \let\hyper@linkend\@empty\citep[#1][#2]{#3}}}
  \newcommandtwoopt{\citetads}[3][][]{\href{http://adsabs.harvard.edu/abs/#3}%
    {\def\hyper@linkstart##1##2{}%
     \let\hyper@linkend\@empty\citet[#1][#2]{#3}}}
  \newcommandtwoopt{\citeyearads}[3][][]%
    {\href{http://adsabs.harvard.edu/abs/#3}
    {\def\hyper@linkstart##1##2{}%
     \let\hyper@linkend\@empty\citeyear[#1][#2]{#3}}}
\makeatother
\usepackage{graphicx}
\usepackage{multirow}
\usepackage{siunitx}  
\usepackage{txfonts}

\begin{document}

\title{The galaxy group merger origin of the Cloverleaf odd radio circle system} 

\author{E.~Bulbul\inst{1}, 
X.~Zhang\inst{1}, 
M.~Kluge\inst{1},
M.~Br\"uggen\inst{2}, 
B. Koribalski\inst{3,4},
A.~Liu\inst{1},
E.~Artis\inst{1}, 
Y.~E.~Bahar\inst{1},
F.~Balzer\inst{1},
C.~Garrel\inst{1},
V.~Ghirardini\inst{1},
N. Malavasi\inst{1}, 
A.~Merloni\inst{1}, 
K.~Nandra\inst{1}, 
M.~E.~Ramos-Ceja\inst{1}, 
J.~S.~Sanders\inst{1}, and
S.~Zelmer\inst{1} 
}
\institute{
Max Planck Institute for Extraterrestrial Physics, Giessenbachstrasse 1, 85748 Garching, Germany 
\and
Hamburg Observatory, University of Hamburg, Gojenbergsweg 112, 21029 Hamburg, Germany
\and
CSIRO Astronomy and Space Science, Australia Telescope National Facility, P.O. Box 76, NSW 1710, Australia 
\and
School of Science, Western Sydney University, Locked Bag 1797, Penrith, NSW 2751, Australia
}

\date{March 2024}
\titlerunning{The Cloverleaf ORC}
\authorrunning{Bulbul et al.}

\abstract{
Odd radio circles (ORCs) are a newly discovered class of extended faint radio sources of unknown origin. We report the first detection of diffuse X-ray gas at the location of a low-redshift ORC (z=0.046) known as Cloverleaf ORC. This observation was performed with the XMM-Newton X-ray telescope. The physical extent of the diffuse X-ray emission corresponds to a region of approximately 230~kpc by 160~kpc, lying perpendicular to the radio emission detected by ASKAP. The X-ray spectrum shows characteristics of thermal multiphase gas with temperatures of $1.10\pm0.08$~keV and $0.22\pm0.01$~keV and a central density of $(4.9\pm0.6)\times10^{-4}$~cm$^{-3}$, indicating that the Cloverleaf ORC resides in a low-mass galaxy group. Using  X-ray observations, with hydrostatic equilibrium and isothermal assumptions, we measure the galaxy group to have a gas mass and a total mass of $(7.7\pm 0.8) \times 10^{11}$ M$_{\rm sun}$ and $2.6\pm0.3\times10^{13}$~$M_{\rm sun}$ within the overdensity radius R$_{500}$. The presence of a high-velocity subgroup identified in optical data, the orientation of the brightest cluster galaxy, the disturbed morphologies of galaxies toward the east of the Cloverleaf ORC, and the irregular morphology of the X-ray emission suggest that this system is undergoing a galaxy group merger. The radio power of the ORC could be explained by the shock reacceleration of fossil cosmic rays generated by a previous episode of black hole activity in the central active galactic nucleus.

} 

\maketitle

\section{Introduction} \label{sec:intro}  
Recent radio surveys, including the Australian SKA Pathfinder (ASKAP),  revealed a novel category of faint radio sources known as odd radio circles (ORCs), which were later confirmed with the Giant Metrewave Radio Telescope (GMRT) and MeerKAT \citep{Norris2021, Koribalski2021, Koribalski2023, Lochner2023}. Only eight ORCs are associated with an optical counterpart, elliptical galaxies with reported stellar masses of approximately $\sim10^{11}$~$M_{\rm sun}$, also known as cosmological or extragalactic ORCs. The redshifts of the galaxies associated with the known ORCs range between 0.05 and 0.6, with a typical ORC size of 300~kpc to 500~kpc. Many of these sources have large ring-like emission, bright and clumpy at the edges, with several irregular emission peaks in their centers. The reported radio powers are several times $10^{23}$~W~Hz$^{-1}$ and have radio brightness of 2--9 mJy at GHz frequencies.

Various physical explanations of the nature of the extragalactic  ORCs are proposed. These include the forward termination shock resulting from past starburst events \citep{Norris2021, Norris2022}, binary supermassive black hole mergers \citep{Koribalski2021, Norris2022}, supernova remnants within the Local Group \citep{Filipovic2022, Omar2022a, Sarbadhicary2023}, tidal disruptions of stars by intermediate-mass or supermassive black holes \citep{Omar2022b}, synchrotron emission from virial shocks around massive galaxies \citep{Yamasaki2023} and synchrotron emission from the historical activity of galactic outflows \citep{Coil2023}, and end-on active galactic nucleus (AGN) jet-inflated bubbles \citep{Lin2024}. Recently, \citet{Dolag2023} postulated that the internal shocks from forming the group-size halos might explain the detection frequency and morphology of the ORC populations by employing nonradiative high-resolution simulations of a Milky Way-mass galactic halo. The halos with the dark matter halo virial mass of $10^{12}-10^{13}$~M$_{\rm sun}$ with stellar masses of several times $10^{11}$~M$_{\rm sun}$ with internal shock Mach number 2.1--2.4 may explain these observed features. However, a direct shock acceleration mechanism of the cosmic-ray electrons fails to reproduce the observed luminosities of the radio emission.

The Cloverleaf ORC, discovered by the Australian SKA Pathfinder (ASKAP), is one of the lowest redshift ORCs known to date \citep{Dolag2023, Koribalski2024}. The optical counterpart of the ORC is the elliptical galaxy GAMA~J113727.46-005047.6 at (RA:$174.3644^\circ$, DEC:$-0.8467^\circ$), hereafter brightest cluster galaxy 1 (BCG1), with a stellar mass of $1.4\pm0.3$~$\times10^{11}M_{\rm sun}$ with a spectroscopic redshift of $0.046399\pm1.2\times10^{-5}$ \citep{Driver2022}. Previously, the Cloverleaf system was cataloged as a galaxy group under the name MZ~07128 in the 2dF survey (2dFGGC) galaxy group catalog in the Sloan Digital Sky Survey \citep{Merchan2002}. The galaxy GAMA~J113727.46-005047.6 (CATAID = 534655)\footnote{https://www.gama-survey.org/dr4/tools/sov.php} is also classified as the central galaxy of a low-mass galaxy group with eight total members using the friends-of-friends method in the complete Galaxy and Mass Assembly~I (GAMA-I) survey galaxy group catalog (G3Cv10) \citep{Robotham2011}, with group ID 200792.

In this work we report the first detection of diffuse X-ray emission in the Cloverleaf ORC using XMM-Newton observations. Combining the observed X-ray, optical, and radio properties of the ORC, we explore the possible nature of these intriguing sources. In Section~\ref{sec:analysis}, we describe the X-ray data reduction and analysis. The results from multi-wavelength data and discussion based on the results are provided in Section~\ref{sec:results}, and finally, we present our conclusions in Section~\ref{sec:conc}.

\begin{figure*}
    \centering
    \includegraphics[width=\textwidth]{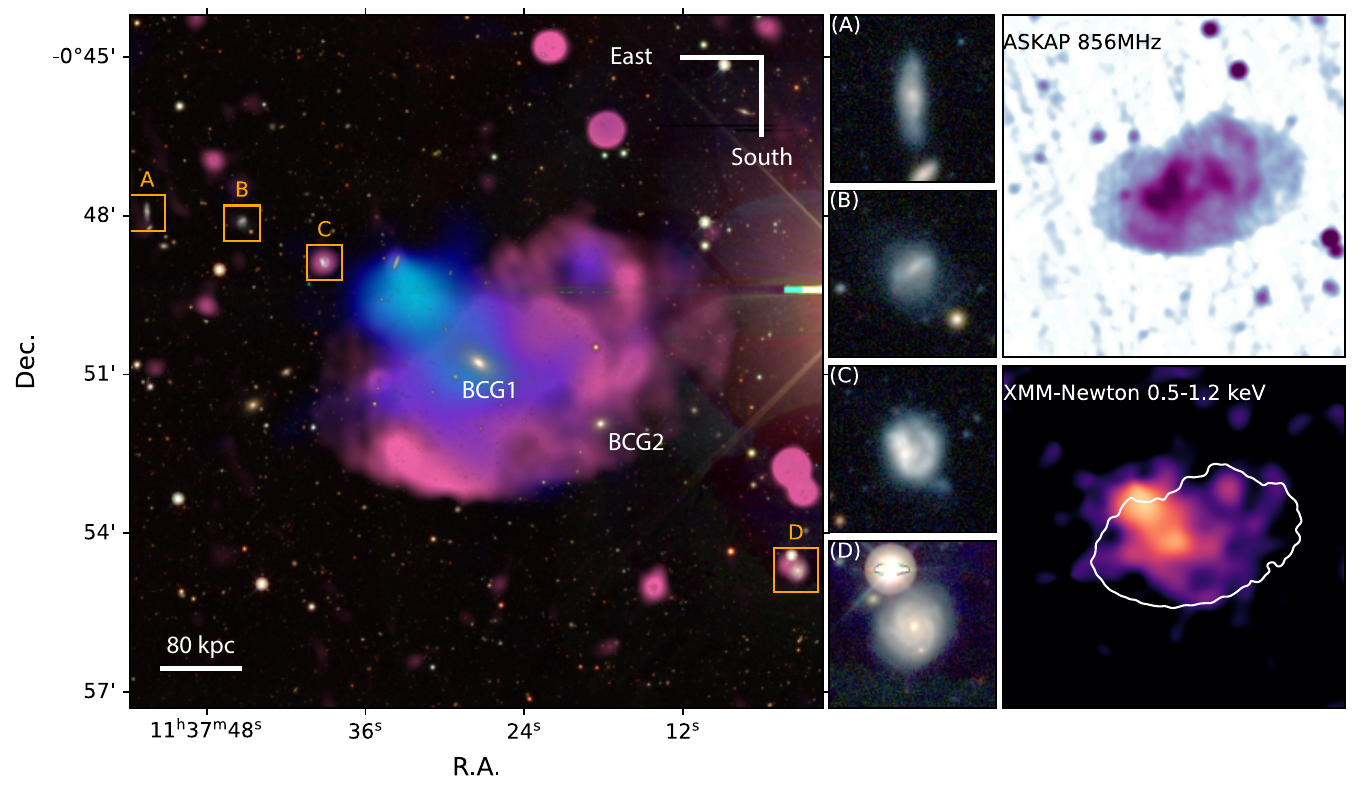}
    \caption{ Multi-wavelength image of the Cloverleaf odd radio circle system.
    \emph{Left:} DESI Legacy Survey DR10 optical (white to yellow), \emph{XMM-Newton} X-ray (blue), and ASKAP radio continuum (red) composite image of the Cloverleaf ORC. 
    \emph{Middle:} Zoomed-in views of three morphologically disturbed galaxies (A, B, C) to the east and morphologically undisturbed galaxy (D) to the west. The BCG1, located at the X-ray and radio emission peak, and another bright elliptical galaxy (BCG2) are also shown. 
    \emph{Top right:} ASKAP 856~MHz image of the Cloverleaf ORC convolved to 20 arcsec resolution. 
    \emph{Bottom right:} EPIC-pn flux image in the 0.5--1.2 keV band. The image is source-removed and smoothed by a $20\arcsec$ Gaussian kernel. Overlaid is the radio ASKAP contour of the Cloverleaf ORC \citep{Koribalski2024}.
    }
    \label{fig:xray_image}
\end{figure*}
\section{X-ray data analysis}
\label{sec:analysis}  

The \emph{XMM-Newton} observations of the Cloverleaf ORC were performed on November 28, 2023, as a Director’s Discretionary Time (DDT) observation 0932390201 (PI: Bulbul). Due to the bright star in the field of view at R.A.=174.23 and Dec.=-0.82, the filter wheels of the two MOS detectors were automatically switched to the CalClosed positions. Therefore, only the data from the PN detector is available for scientific analysis. We used Scientific Analysis System (SAS) 21.0.0 software for data reduction, which follows the process described in \citet{Zhang2023}. After filtering the flare time intervals, the clean exposure time is 21.6~ks. 

\subsection{Imaging analysis}
We adopt the 0.5--1.2~keV band for imaging analysis. Point sources were detected using the \texttt{ewavelet} algorithm with parameters \texttt{minscale=1}, \texttt{maxscale=16}, and \texttt{threshold=5}. The resulting XMM-Newton X-ray, ASKAP, and composite X-ray, radio, and optical images from the DESI Legacy Data Release (DR10) data are shown in Fig.~\ref{fig:xray_image}.

We extracted an azimuthally averaged surface brightness profile using the 0.5--1.2~keV images around the centroid of the Cloverleaf ORC. The centroid of the system, which is assumed to be the middle of the two X-ray peaks, was adopted as the center of the extraction region. We fit the surface brightness profile using a $\beta$-profile density model \citep{Cavaliere1976},

\begin{equation}
    n_\mathrm{e}(r) = n_\mathrm{e,0}\left[1+\left(\frac{r}{r_\mathrm{c}}\right)\right]^{-3\beta/2},
\end{equation}

\noindent where $n_\mathrm{e,0}$ is the central electron density, {\bf $r_\mathrm{c}$ is the core radius}, and the parameter $\beta$ defines the slope of the density profile.

\subsection{Spectral analysis}
We selected \texttt{FLAG==0} events for spectral analysis within the 3\arcmin\ radius region surrounding the centroid of the two X-ray peaks shown in Fig.~\ref{fig:xray_image}. The background components consist of celestial X-ray photons and \ac{nxb} components. The former includes foreground emission from the \ac{gh} and \ac{cxb} as described in \citet{Bulbul2012}. In addition to the \xmm spectra, we extracted the combined spectrum (TM8) of the public eROSITA All-Sky Survey (eRASS1) observations in a region between 0.5--1~degree annulus centered on the Cloverleaf ORC by \texttt{srctool} in the eROSITA Science Analysis Software System (eSASS) software to constrain the foreground components and the contribution from the Milky Way halo to the background emission. The detected eRASS1 point and extended sources are masked when extracting the eRASS1 spectrum using the public catalogs \citep{Bulbul2024, Merloni2024, Kluge2024}. 

We used the X-ray spectral fitting package (XSPEC) to co-fit the XMM-Newton and eROSITA spectra. All spectra were optimally binned using \texttt{ftgrouppha} before loading into XSPEC \citep{Kaastra2016}. The detailed configurations of the background model components are described in Appendix~\ref{app:spec}. The four spectra were fit simultaneously to constrain the temperature of the diffuse X-ray emission and model the foreground and the particle background.
We extracted a global spectrum using a $3\arcmin$ aperture, including the northern and southern X-ray peaks. The low signal-to-noise PN observations did not allow a detailed study of the metal and temperature distribution of the system. Therefore, we froze the abundance to be $0.3Z_{\rm sun}$ for the rest of the spectral fits as typically assumed in galaxy clusters and groups \citep[e.g.,][]{Bulbul2016, Mernier2018, Liu2020}.

\section{Results and discussion}
\label{sec:results}  

The radio emission in the 856~MHz ASKAP image shows an elliptical boundary with a major axis of $4\arcmin$ (227 kpc) to a minor axis of $2.8\arcmin$ (137~kpc) with an inclination angle of $\sim15^\circ$ as plotted in the right panel of Fig.~\ref{fig:xray_image}. The internal radio morphology clearly shows two peaks east of the ORC, 1~arcmin (55~kpc) apart from each other, with a shell-like emission surrounding the radio peaks. One of the two peaks is located at the position of the central galaxy. The detailed radio study of this ORC will be presented in \citet{Koribalski2024}.

\subsection{Properties of the diffuse gas}
\label{sec:IGrM}

The new XMM-Newton observations show a clear diffuse X-ray emission at the location of the Cloverleaf ORC. The extended emission lies in the northeast direction perpendicular to the radio emission with a region size similar to a region approximately the size of an ellipse with axes of $\sim1.9\arcmin$ (100~kpc) to $3.3\arcmin$ (180~kpc). The X-ray morphology of diffuse gas shows two clear peaks; the central galaxy is located at the center of the southern peak, which also marks the location of the brightest radio emission. In contrast, the X-ray peak in the northeast direction is $1.8\arcmin$~(98~kpc) from the southern peak and is not associated with a galaxy. The centroid of the X-ray emission (i.e., the center of the two X-ray peaks) is 37~kpc from the center of the two radio peaks and 80~kpc from the center of the radio ellipse. The southern peak also shows clear X-ray detection of an AGN in the center. The lack of the optical counterpart at the center of the northeastern X-ray peak and the disturbed morphology of the intragroup medium (IGrM) indicate that this system is in a nonrelaxed dynamical state.

\begin{figure}
    \centering
    \includegraphics[width=0.5\textwidth]{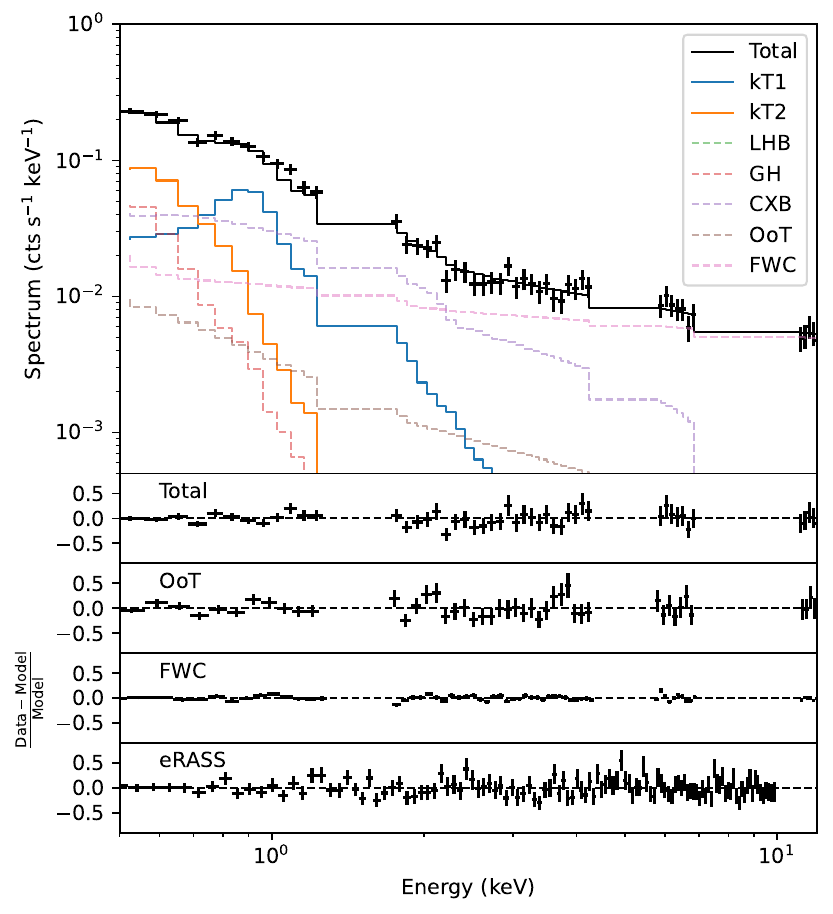}
    \caption{XMM-Newton spectrum of the Cloverleaf ORC is shown in black in the top panel. The best-fit two-temperature IGrM model is shown as blue and orange, while the foreground and background models are plotted as dashed lines. The bottom panels indicate good fits after the background modeling, including NXB with the foreground modeling of the eRASS1 data.}
    \label{fig:spec}
\end{figure}

The X-ray spectroscopy provides measurements of the ambient gas temperature of the intragroup medium. The spectral fits adopting a single temperature scenario and free metal abundance yield a best-fit model with zero metallicity, suggesting a strong Fe-bias and the existence of a multi-temperature phase in the system \citep{Simionescu2009}. The best-fit temperature of the model is $k_\mathrm{B}T=0.74\pm0.04$~keV in a single-temperature scenario with a C-stat value of 358.2/283. The two-temperature model significantly improves the spectral fit with components with $k_\mathrm{B}T_1=1.10\pm0.08$~keV and $k_\mathrm{B}T_2=0.23\pm0.02$~keV (C-stat value of 277.9/281). 
The X-ray source, background spectra, and the two-temperature model fits are shown in Fig. \ref{fig:spec}. We also fit the individual spectra of the northeastern and southern X-ray emission. In a single-temperature model, the temperature of the northeastern X-ray peak is $0.75\pm0.05$~keV, and the southern peak has a slightly lower temperature of $0.31\pm0.03$~keV.

By fitting the azimuthally averaged X-ray surface brightness profile obtained from the XMM-Newton imaging analysis, we measure the central electron density to be $n_\mathrm{e,0}=(4.9\pm0.6)\times10^{-4}$~cm$^{-3}$. To measure the total mass at the physically meaningful overdensity radius, R$_{500}$, defined as the radius within which the average density of the matter is 500 times the critical density of the Universe at the group's redshift, we use an isothermal IGrM average temperature of $0.74\pm0.04$~keV \citep[see][for the details of the method]{Bulbul2012}. Under the assumption of hydrostatic equilibrium, we find that the total mass is $(2.6\pm0.3) \times 10^{13}$~M$_{\rm sun}$ at R$_{500}$ of $446 \pm 40$~kpc. Using the relation R$_{500}= 0.659$~R$_{200}$ \citep[see][]{Bulbul2016}, we find that the virial radius is $\sim$680~kpc. We note that the assumption of an isothermal IGrM here may underestimate the M$_{500}$ uncertainty, and the current uncertainty reflects the precision of the measured electron density profile. Integrating the gas density over a volume encapsulated by the overdensity radius R$_{500}$ gives a gas mass of $(7.7 \pm 0.8) \times 10^{11}$~M$_{\rm sun}$. The fraction of gas mass to total mass yields a gas mass fraction of $0.03\pm0.003$, lower than the typical X-ray selected galaxy groups \citep{Bulbul2024, Bahar2024}.

In the 0.5--2.0~keV band, the flux and luminosity of the ORC are $(2.3\pm0.2)\times10^{-13}$~erg~s$^{-1}$~cm$^{-2}$ and $(1.3\pm0.1)\times10^{42}$~erg~s$^{-1}$ within R$_{500}$, respectively. We compare our luminosity and temperature measurements with the scaling relations of the eFEDS sample, including a larger number of galaxy groups, and find that the X-ray properties of the Cloverleaf system are consistent with the L$_{\rm X}-$~kT relations of the eFEDS sample given the large scatter in these low-mass, low-IGrM temperature regimes \citep{Bahar2022}.

\subsection{Dynamical state and the origin of the ORC}

\begin{figure}
    \centering
    \includegraphics[width=0.5\textwidth]{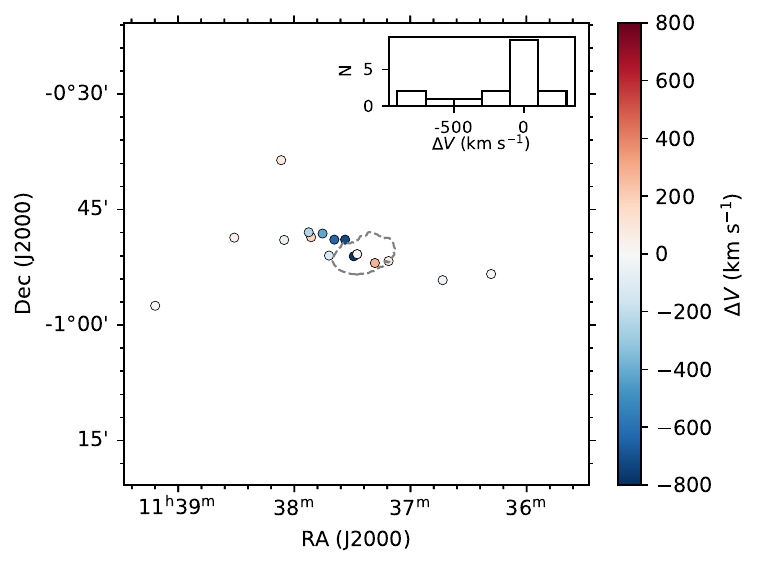}
    \caption{Line-of-sight velocity structures of galaxies with $z_\mathrm{spec}$ data within $30\arcmin$ angular distance. The dashed contour is the boundary of the radio emission. The inset is the histogram of $\Delta V$.}
    \label{fig:delta-v}
\end{figure}

The catalogs in \citealt{Kluge2024} (see Appendix~D) provide the photometric and spectroscopic redshifts of the galaxies in the literature in the western Galactic hemisphere. We search for galaxies with $|\Delta V|<1000$~km~s$^{-1}$ ($5\times\sigma_\mathrm{V}$ of the group), around the Cloverleaf ORC. The location and their velocity dispersion are shown in Fig.~\ref{fig:delta-v}. The projected spatial distribution of galaxies within this velocity range is elongated in an east--west orientation. A substructure on the northeast boundary of the radio emission has a velocity discrepancy with $>600$~km~s$^{-1}$ from the central galaxy. This substructure is also identified as an individual galaxy group (ID~201611) in the catalog of G3Cv10 \citep{Robotham2011}. 

By inspecting the morphology of individual galaxies in the DESI Legacy DR10 data
\citep{Dey2019},\footnote{\url{https://www.legacysurvey.org}} we find that three galaxies in the northeast of the central galaxy (A, B, and C in Figure \ref{fig:xray_image}), which are possible members of the subgroup at similar redshifts, show disturbed morphologies. Galaxy A has a warped disk slightly stretched in the southwest direction, galaxy B has elongated diffuse stellar emission toward the south, and galaxy C has a disturbed disk with a blue clump in the southwest unrelated to the disk structure. This clump may indicate triggered star formation due to ram-pressure stripping as it is directed toward the X-ray emission. Conversely, galaxy D ($z_\mathrm{phot}=0.054\pm0.010$) shows a relaxed disk morphology. In our merger scenario, this galaxy is still in-falling. It has not yet undergone tidal interactions with the central galaxy, contrary to galaxies A, B, and C, with post-merger distortions in their morphology. 

BCG1 is located at the center of the  X-ray and radio peaks, while another bright elliptical galaxy (BCG2 in Fig.~\ref{fig:xray_image}) is detected to the southwest of BCG1. It is likely that BCG2, which might be associated with the northeastern X-ray group, has already passed through the southern group due to the merger while lagging behind the highly collisional X-ray-emitting gas.

\subsection{Comparison to other diffuse radio sources in groups and clusters}
If the radio emission in the Cloverleaf ORC is caused by shock acceleration from merger shocks, it is worth comparing its properties to radio relics in galaxy clusters \citep[e.g.,][]{Koribalski2023}. Radio relics have powers of $\sim 10^{25}$~W/Hz at 1.4~GHz, and their power is a strong function of the cluster mass \citep{Jones2023}. Their coincidence with X-ray-detected shock waves has suggested that they are caused by merger shocks where cosmic-ray electrons are accelerated with fairly high efficiencies \citep{2020A&A...634A..64B}. These shocks are usually found well within the cluster virial radius and have a moderate sonic Mach number of 2-3, unlike the cluster accretion shocks that sit outside the virial radius and have much higher Mach numbers.

\begin{figure}
    \centering
    \includegraphics[width=0.5\textwidth]{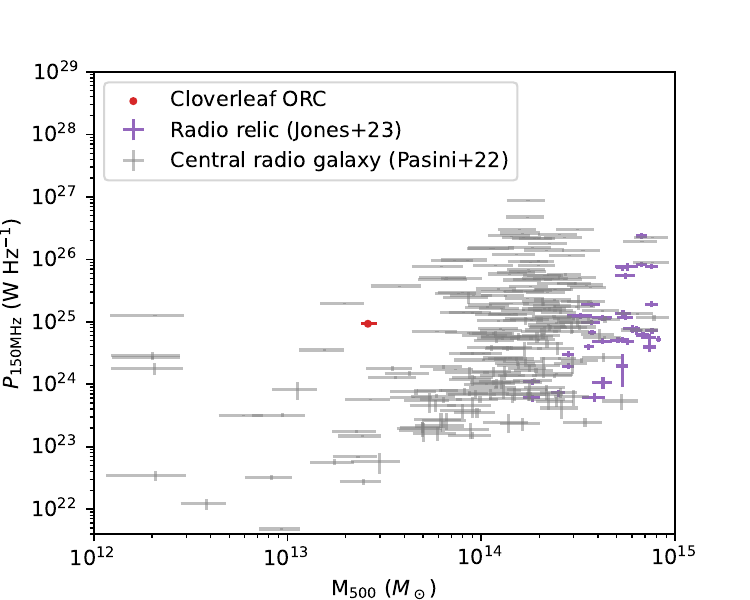}
    \caption{Diagram of radio power vs. halo mass for the samples of radio relics \citep{Jones2023} and central radio galaxies \citep{Pasini2022} observed by LOFAR and the Cloverleaf ORC. The power of the Cloverleaf ORC is converted to the frequency of 150 MHz by assuming a $\alpha=1$ spectral index. The error bar of the Cloverleaf ORC radio power is within the mark. }
    \label{fig:power-mass}
\end{figure}

In Fig.~\ref{fig:power-mass}, we plot the radio luminosities of radio relics versus their mass. Here, we use the $M_{500}$ of the Cloverleaf group measured in Sect.~\ref{sec:IGrM}. We also scale the radio power of the Cloverleaf ORC from the 856~MHz frequency to the LOFAR high-band antenna central frequency of 150 MHz by assuming $\alpha=1$, which is the typical spectral index for radio relics in clusters \citep{vanWeeren2019}. We find a mass difference of almost two orders of magnitude between the Cloverleaf ORC and known radio relics. 

Radio relics are megaparsec-sized objects seen nearly edge-on; thus, radio power should decrease in smaller objects. This was confirmed by \citet{Dolag2023}, who found that the high radio power observed in ORCs cannot be reproduced in numerical simulations. Reacceleration of fossil cosmic rays (CRs) could help reproduce the high observed radio power, as recently discussed by \citet{2024arXiv240209708S}. These fossil cosmic rays could be injected by a central radio galaxy (e.g., powered by the AGN found in the Cloverleaf ORC). Then, the passage of a shock wave can sweep up the radio lobe and re-accelerate the plasma, creating the ORC. \citet{2024arXiv240209708S} discuss the combinations of viewing angles, angles between shock and radio lobe axis, as well as the age of the lobes that would lead to the observed ORC properties. This formation scenario also naturally explains the rarity of ORCs since only fairly narrow ranges in lobe ages and geometric factors lead to observable ORCs. Future simulations, including the role of magnetic fields, will have to explore this possibility. This scenario might also be validated by radio observations at very low radio frequencies that may detect the emission of remnant lobes.

The high radio power also suggests high magnetic field strengths, and, for reference, we can estimate the equipartition magnetic field. 
 If the radio emission is homogeneous within the volume of the Cloverleaf ORC, the source depth is the average length of the long and short axes, which is 360~kpc. Meanwhile, we use the radio surface brightness close to the edge for a lower limit estimate. We estimate a minimum energy density contained in \acp{cr} of $u_\mathrm{min,CR+B}=3.2\times10^{-14}$~erg~cm$^{-3}$ \citet{Govoni2004}, adopting $\alpha=1$. The corresponding revised magnetic field strength at equipartition is therefore $B_\mathrm{eq}^{'}=1.2$~$\mu$G by assuming $\gamma_\mathrm{min}=100$ \citep{Brunetti1997,Beck2005}. Both $u_\mathrm{min}$ and $B_\mathrm{eq}^{'}$ are higher if we adopt steeper spectral indices. The estimated $1.2\,\mu$G equipartition magnetic field strength results in a magnetic pressure of $P_\mathrm{B}=B^2/8\pi\sim3.6\times10^{-5}$~keV~cm$^{-3}$. 
For comparison, the thermal pressure in the hot IGrM can be calculated as $P_\mathrm{th}=nk_\mathrm{B}T\sim6.9\times10^{-4}$~keV~cm$^{-3}$, taking the central gas density as an upper limit. The ratio of the thermal and magnetic pressures, $\beta_\mathrm{PL}\equiv P_\mathrm{th}/P_\mathrm{B}$, is $\beta_\mathrm{PL}\sim20$ compared to $>100$ in typical galaxy clusters \citep{Donnert2018}. However, we also note that in galaxy clusters, magnetic field strengths are around 1/20th of the equipartition field \citep{2019LRCA....5....2B}.

\section{Conclusions}
\label{sec:conc}  
This work presented the first detection of diffuse X-ray emission from a recently discovered Cloverleaf odd radio circle. The extended X-ray source with an extent of 100~kpc by 180~kpc lies in the northeast direction perpendicular to the radio emission. The total 0.5-2 keV luminosity is $(1.3\pm 0.1) \times 10^{42}$~erg~s$^{-1}$. The X-ray emission has two clear peaks. The southern X-ray peak has a counterpart, an elliptical central galaxy (BCG1), which also marks the location of the brightest radio emission. The northeastern peak does not have an optical counterpart or correspond to any radio peak. The offset between the centroid of the X-ray emission and radio emission is 80~kpc.

The X-ray emission has a thermal nature, where the average density of the gas is $(7.0\pm1.8) \times 10^{-5}$~cm$^{-3}$ within R$_{500}$, which is lower than typically observed in galaxy groups \citep{Bahar2024}. We measure the central electron density to be $(4.9 \pm 0.6)\times 10^{-4}$~cm$^{-3}$. The temperature measurements of the gas indicate that it is in a multi-phase state with components of $1.10\pm0.08$~keV and $0.23\pm0.02$~keV. The corresponding gas and total masses are $(7.7\pm0.8)\times 10^{11}$~M$_{\rm sun}$ and $(2.6\pm0.3) \times 10^{13}$~M$_{\rm sun}$ within the overdensity radius of R$_{500}$ of $446\pm40$~kpc, inferred from the X-ray observations under the hydrostatic equilibrium and isothermal IGrM assumptions. The flux and luminosity of the ORC are $(2.3\pm0.2)\times10^{-13}$~erg~s$^{-1}$~cm$^{-2}$ and $(1.3\pm0.1)\times10^{42}$~erg~s$^{-1}$ measured within the same radius. The properties of the diffuse X-ray thermal gas indicate that the Cloverleaf ORC lies in an intra-group environment.

The disturbed morphology of the intra-group medium, the high-velocity subgroup identified at optical wavelengths, and the disturbed morphologies of galaxies toward the east around the Cloverleaf ORC suggest that this system is undergoing a merger activity. The absence of the central galaxy of the northeastern gas component may indicate that the collisionless member galaxies have already left behind the highly collisional intragroup medium. Such large offsets between the galaxy and X-ray gas are commonly observed in major galaxy cluster mergers \citep{Markevitch2004}.

If the elliptical morphology of radio emission has a shock origin, comparing the radius of radio emission of the ORC ($\sim200$~kpc) and the virial radius of the galaxy group $\sim680$~kpc, the merger shock is the most likely explanation of the observed multiwavelength properties of this object. With eight known ORCs, any explanation of their origin must explain why they are rare. \citet{Dolag2023} argue that the galaxy mergers that produce ORCs must lead to a dramatic increase in the halo mass (by a factor of $\sim 3$), which can explain why so few are seen. Moreover, there may be further constraints on the angle between the main merger axis and the line of sight, making them rarer still.

Our results indicate that, for the Cloverleaf ORC, merger shocks in galaxy group size halos are the most likely origin for the radio emission. To explain its high radio luminosity, a scenario that involves remnant radio lobes of a supermassive black hole, as discussed in \citet{2024arXiv240209708S}, is plausible.

\begin{acknowledgements}

The authors thank the referee for helpful and constructive comments on the draft. E. Bulbul, X. Zhang, A. Liu, V. Ghirardini, C. Garrel, and S. Zelmer, acknowledge financial support from the European Research Council (ERC) Consolidator Grant under the European Union’s Horizon 2020 research and innovation program (grant agreement CoG DarkQuest No 101002585). 
MB acknowledges funding by the Deutsche Forschungsgemeinschaft under Germany's Excellence Strategy -- EXC 2121 ``Quantum Universe'' --  390833306 and FOR 5195.
N. Malavasi acknowledges funding by the European Union through a Marie Skłodowska-Curie Action Postdoctoral Fellowship (Grant Agreement: 101061448, project: MEMORY)."
\\
This work is based on observations obtained with XMM-Newton, an ESA science mission with instruments and contributions directly funded by ESA Member States and NASA'.
\\
This work uses public data from eROSITA, the soft X-ray instrument aboard SRG, a joint Russian-German science mission supported by the Russian Space Agency (Roskosmos), in the interests of the Russian Academy of Sciences represented by its Space Research Institute (IKI), and the Deutsches Zentrum f{\"{u}}r Luft und Raumfahrt (DLR). The SRG spacecraft was built by Lavochkin Association (NPOL) and its subcontractors and is operated by NPOL with support from the Max Planck Institute for Extraterrestrial Physics (MPE).

The development and construction of the eROSITA X-ray instrument was led by MPE, with contributions from the Dr. Karl Remeis Observatory Bamberg \& ECAP (FAU Erlangen-Nuernberg), the University of Hamburg Observatory, the Leibniz Institute for Astrophysics Potsdam (AIP), and the Institute for Astronomy and Astrophysics of the University of T{\"{u}}bingen, with the support of DLR and the Max Planck Society. The Argelander Institute for Astronomy of the University of Bonn and the Ludwig Maximilians Universit{\"{a}}t Munich also participated in the science preparation for eROSITA.

The eROSITA data shown here were processed using the \esass software system developed by the German eROSITA consortium.

\\

The Legacy Surveys consist of three individual and complementary projects: the Dark Energy Camera Legacy Survey (DECaLS; Proposal ID \#2014B-0404; PIs: David Schlegel and Arjun Dey), the Beijing-Arizona Sky Survey (BASS; NOAO Prop. ID \#2015A-0801; PIs: Zhou Xu and Xiaohui Fan), and the Mayall z-band Legacy Survey (MzLS; Prop. ID \#2016A-0453; PI: Arjun Dey). DECaLS, BASS, and MzLS together include data obtained, respectively, at the Blanco telescope, Cerro Tololo Inter-American Observatory, NSF’s NOIRLab; the Bok telescope, Steward Observatory, University of Arizona; and the Mayall telescope, Kitt Peak National Observatory, NOIRLab. Pipeline processing and analyses of the data were supported by NOIRLab and the Lawrence Berkeley National Laboratory (LBNL). The Legacy Surveys project is honored to be permitted to conduct astronomical research on Iolkam Du’ag (Kitt Peak), a mountain with particular significance to the Tohono O’odham Nation.

\\

This scientific work uses data obtained from Inyarrimanha Ilgari Bundara / the Murchison Radio-astronomy Observatory. We acknowledge the Wajarri Yamaji People as the Traditional Owners and native title holders of the Observatory site. CSIRO’s ASKAP radio telescope is part of the Australia Telescope National Facility (https://ror.org/05qajvd42). Operation of ASKAP is funded by the Australian Government with support from the National Collaborative Research Infrastructure Strategy. ASKAP uses the resources of the Pawsey Supercomputing Research Centre. Establishment of ASKAP, Inyarrimanha Ilgari Bundara, the CSIRO Murchison Radio-astronomy Observatory and the Pawsey Supercomputing Research Centre are initiatives of the Australian Government, with support from the Government of Western Australia and the Science and Industry Endowment Fund.

\\

\end{acknowledgements}

\bibliography{references}{}

\appendix

\section{Assumed cosmology}\label{app:spec}

In this work we adopt a $\Lambda$-cold-dark-matter cosmology with parameters $H_0=70$~km~s$^{-1}$~Mpc$^{-1}$, $\Omega_\mathrm{m}=0.3$, and $\Omega_\Lambda=0.7$. At $z=0.0464$, one arcsecond corresponds to a physical scale of 0.91~kpc. 

\section{Imaging analysis}\label{app:img}
For extracting the image, we selected \texttt{(FLAG \& 0xfb0825)!=0 \&\& PATTERN<=4} events. 
We adopted the cooling function in the $[0.5,1.2]\times(1+z)$~keV band at the system's average temperature to convert the density to projected surface brightness.

\section{Spectral analysis and models}\label{app:spec}

Energy ranges, including fluorescent instrumental lines in the energy bands of 1.3--1.7~keV, 4.4--5.7~keV, and 7--11~keV, were ignored in the PN fittings.  We used C-stat to calculate the fitting statistics \citep{Cash1979} and adopted the solar abundance table from \citet{Asplund2009}. 

The components in different spectra are listed in Table \ref{tab:spec_comp}. Among them, the components of celestial X-ray photons are convolved by the \ac{rmf} and folded by the \ac{arf} for model calculation; components of particle background and \ac{oot} are only convolved by the \ac{rmf} for model calculation. The CXB includes 6.3\% of \ac{oot} events and particle backgrounds whose spectra were generated from \ac{oot} event files and the stacked \ac{fwc} event file,\footnote{https://www.cosmos.esa.int/web/xmm-newton/filter-closed} respectively.

The detailed XSPEC spectral model definition of each component is listed in Table \ref{tab:spec_model}. The redshifts of the source APEC models are fixed to 0.046; the hydrogen column density of the \texttt{tbabs} models are fixed to $2.3\times10^{20}$, which includes the contribution from both neutral hydrogen and molecular hydrogen \citep{Willingale2013}. The redshifts of both the \ac{lhb} and \ac{gh} APEC models are fixed to zero. The abundance of the \ac{lhb} APEC is fixed to 1.0 while that of the \ac{gh} APEC model is set to be free, given the recent $\sim0.06$~$Z_{\rm sun}$ measurement reported by \citet{Ponti2023}. Because we already ignored the energy ranges of instrumental lines, we modeled the pn \ac{fwc} spectrum using a \texttt{bkg2pow} model. We used a free constant parameter to scale the pn \ac{fwc} normalization to the particle background level in the observed spectrum. The pn \ac{oot} spectrum is a highly smoothed source spectrum, which can be well fitted empirically using a \texttt{bkg2pow} model (see the fitting residual of the \ac{oot} spectrum in Fig. \ref{fig:spec}). The \ac{fwc} spectral model was provided by \citet{Yeung2023}, which is publicly available on the eROSITA DR1 site.\footnote{\url{https://erosita.mpe.mpg.de/dr1/AllSkySurveyData_dr1/FWC_dr1/}} 

\begin{table*}[]
\centering

    \caption{Spectral components and the corresponding responses for individual spectra.}
    \label{tab:spec_comp}
    
    \begin{tabular}{ccc}
        \hline\hline
        Spectrum & \multicolumn{2}{c}{Components} \\
        \cline{2-3}
        & w/ RMF \& ARF & w/ RMF \\ 
        \hline
         EPIC-pn &  SRC + FG + CXB-XMM & $0.063\times$OOT + constant$\times$FWC-pn\\
         EPIC-pn OoT & - & OOT\\
         EPIC-pn FWC & - & FWC-pn \\
         eRASS:1 TM8 local background& FG + CXB-eRASS & FWC-TM8\\
        \hline
    \end{tabular}
    
\end{table*}

\begin{table*}[]
    \caption{Spectral model for each component}\label{tab:spec_model}
    \centering
    \begin{tabular}{cc}
        \hline\hline
         Component &  Models\\
         \hline
         SRC & \texttt{tbabs*(apec+apec)}\\
         FG & \texttt{apec+tbabs*apec} \\
         CXB-XMM & \texttt{powerlaw}\\
         CXB-eRASS & \texttt{powerlaw}\\
         OOT & \texttt{bkn2pow}\\
         FWC-pn & \texttt{bkn2pow}\\
         FWC-TM8 & Model combination by \citet{Yeung2023}\\
         \hline
    \end{tabular}
    
\end{table*}

\section{Averaged surface brightness profile}
\begin{figure}
    \centering
    \includegraphics[width=0.5\textwidth]{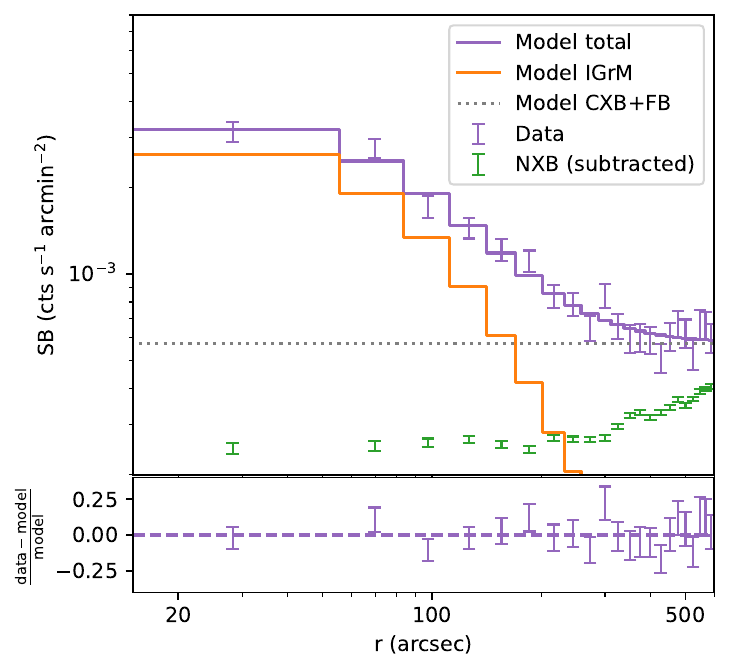}
    \caption{Azimuthally averaged surface brightness profiles (purple),  subtracted \ac{nxb} (green), and best-fit model profile (purple step). The models of the individual components, i.e., the IGrM and X-ray foreground and background, are plotted as orange steps and dotted gray lines, respectively.}
    \label{fig:sb}
\end{figure}

We plot the azimuthally averaged surface brightness profile as well as the best-fit models in Fig. \ref{fig:sb}.

\section{Morphology of Galaxy A}

\begin{figure}
    \centering
    \includegraphics[width=0.45\linewidth]{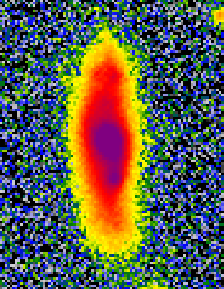}
    \includegraphics[width=0.45\linewidth]{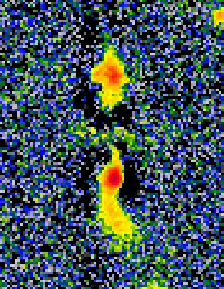}
    \caption{Cutout of the Legacy Surveys $g$-band image around galaxy A (left) and residual after subtracting an isophotal model. North is up; east is left.}
    \label{fig:galAmorph}
\end{figure}

We note an asymmetric feature in the stellar light of galaxy A in Figure \ref{fig:galAmorph}, left panel. The disk is slightly warped. The residual image, after subtracting an isophotal model, enhances this feature in the south by revealing an extension toward the southwest. The isophotal model was created following the procedure described in \cite{Kluge2023}.

\end{document}